\documentclass{article}
\usepackage[utf8]{inputenc}

\usepackage{listings}  

\usepackage{float} 

\usepackage{algorithm} 
\usepackage{algpseudocode}

\usepackage{hyperref}
\usepackage{todonotes}

\title{Transformer Models for Type Inference in the Simply Typed
  Lambda Calculus: A Case Study in Deep Learning for Code}

\author{Brando Miranda \footnote{Also affiliated with the University
    of Illinois at Urbana-Champaign.}, Nathan Fulton, Avi Shinnar, \\
  Vasily Pestun\footnote{Also affiliated with IHES.}, and Barry Trager
  \\
  IBM Research
}

\date{August 2021}

\begin{document}

\maketitle


\begin{abstract}
  Despite a growing body of work at the intersection of deep learning
  and formal languages, there has been relatively little systematic
  exploration of transformer models for reasoning about typed lambda
  calculi.  This is an interesting area of inquiry for two reasons.
  First, typed lambda calculi are the lingua franc of programming
  languages. A set of heuristics that relate various typed lambda
  calculi to effective neural architectures would provide a systematic
  method for mapping language features (e.g., polymorphism, subtyping,
  inheritance, etc.) to architecture choices.  Second, transformer
  models are widely used in deep learning architectures applied to
  code, but the design and hyperparameter space for them is large and
  relatively unexplored in programming language applications.
  Therefore, we suggest a benchmark that allows us to explore exactly
  this through perhaps the simplest and most fundamental property of a
  programming language: the relationship between terms and types.
  Consequently, we begin this inquiry of transformer architectures for
  typed lambda calculi by exploring the effect of transformer warm-up
  and optimizer selection in the task of type inference: i.e.,
  predicting the types of lambda calculus terms using only
  transformers. We find that the optimization landscape is difficult
  even in this simple setting.  One particular experimental finding is
  that optimization by Adafactor converges much faster compared to the
  optimization by Adam and RAdam. We conjecture that such different
  performance of optimizers might be related to the difficulties of
  generalization over formally generated dataset.
\end{abstract}

\section{Introduction}

The success of deep learning in computer vision \cite{alexnet,
  resnet}, natural language processing \cite{bert, gpt3}, and
especially game playing \cite{alphago, atari1, EfficientZero}
motivates a growing body of work on deep learning for programming
tasks.  Work in the intersection of deep learning and programming
languages includes the use of neural networks to solve classical
problems in programming language research (neural guided proof search
\cite{proof_search}, neural deductive program synthesis
\cite{neural_deductive}, and neural inductive program synthesis
\cite{inductive_synthesis}), the use of programming languages to
address the weaknesses of deep learning in canonical application
domains, and the combination of programming languages with deep
learning to solve novel tasks (e.g., NLP-to-code \cite{codex}).

Despite a growing body of work at the intersection of deep learning
and formal languages, there has been relatively little systematic
exploration of transformer models for reasoning about typed lambda
calculi.  This is an interesting area of inquiry for two reasons.
First, typed lambda calculi are a lingua franc of programming
languages.  A set of heuristics that relate various typed lambda
calculi to effective neural architectures would provide a systematic
method for mapping language features (e.g., polymorphism, subtyping,
inheritance, etc.) to architecture choices.  Second, transformer
models are widely used in deep learning architectures applied to code,
but the design space and hyperparameter space for transformers is
truly massive and relatively unexplored in programming language
applications.

This paper begins an inquiry of transformer architectures for simply
typed lambda calculi by exploring the effect of transformer warm-up
\cite{trans_tutorial, Vaswani2017, radam,
  on_the_adequacy_of_untuned_warmup} and optimizer selection for
learning to do type inference.  This benchmark allows us to explore,
in a simple setting,  ability of transformers to reason about
perhaps the simplest and most fundamental property of a programming
language: the relationship between terms and types.

\textbf{Our contributions are the following}:

\begin{itemize}
\item We explore a full spectrum of hyperparameters for a fixed
  transformer architecture.  
  We find that the \emph{learning rate} but not the warm-up steps, is the most important hyperparameter.
  This finding is significant due to the strong emphasis in the
  current literature on the importance of a warm-up schedule for the
  learning rate \cite{trans_tutorial, Vaswani2017, radam, on_the_adequacy_of_untuned_warmup, adafactor}.

\item We show that, despite the simplicity of the task,
    extensive hyperparameter search was needed to achieve acceptable
    performance.
    
  \item We find that the Adafactor optimizer \cite{adafactor}
    consistently achieves zero train error and zero validation error
    on our task without any hyperparameters tuning.  This holds both
    when using complicated path embeddings of formulas (analogous to a
    character embedding for natural language) and when using depth
    embeddings \cite{treegen}.


    From the impressive stability Adafactor brought to our training we conjecture that 
    the difficulty of training transformer-based models for
    learning formal rules -- for example first observered  with ``Grokking" by previous work
    \cite{grokking} -- might be avoidable by using the Adafactor
    optimizer \cite{adafactor}.
    Therefore, we suggest practitioners and
    researchers to try Adafactor first.
\end{itemize}

\section{Background}

This section reviews transformers and the simply typed lambda calculus.

\subsection{Transformers}

The transformer model has had a tremendous impact in natural language
processing (NLP).  
It is able to generate human-like sounding language with
models like the GPT-n series \cite{gpt3}.  
In addition, the transformer model has been used to generate code with GPT-3
\cite{gpt3}, GitHub copilot \cite{codex}, TreeGen \cite{treegen},
GPT-f \cite{gptf}, skip-trees \cite{skiptree} and more.  
A common way it's trained for NLP is by
transforming the natural language sentences into a sequence of tokens
from a closed vocabulary, and then feeding large amounts of data  (e.g., millions of tokens) to a transformer model with an order of millions of parameters \cite{Vaswani2017}.  
In formal language generation, transformers are
typically trained to generate code from natural language \cite{codex,
  treegen} or directly on the formal languages \cite{gptf, skiptree,
  coqgym}.

\subsection{The Simply Typed Lambda Calculus} \label{lambda_calc_background}

The simply typed lambda calculus is a programming language model that
captures the essence of function definitions and function
applications. 
The language has both terms and types.
Note that terms are also called programs, but we will use the word term in our paper.

A type is either an elementary type denoted by a letter from a prefixed based alphabet $t \in T$,
or a type of a function denoted by an arrow from type to type:
  
\begin{equation}
  \label{eq:grammartype}
  \tau ::= t ~|~ \tau \rightarrow \tau
\end{equation}

Terms of the simply typed lambda calculus include variables (ranged
over alphabet of variables $x \in X$), function definitions, and
function applications:

\[
e ::= x ~|~ \lambda x: \tau. e ~|~ e_1~e_2
\]
The expression $\lambda x: \tau. e$ defines a function that takes one
argument called $x$ with type $\tau$; $e$ is the body of the function
(which may, of course, mention the argument $x$).  The expression
$e_1~e_2$ calls the function $e_1$ with the argument $e_2$.

The language can be enriched with base types and base variables.  For
example, consider the language extended with a {\tt nat} base type and
base variables ${\tt 0, 1, 2, \dots}$.  The following expression
defines a function that takes any number as input and returns the
number {\tt 1}, and then applies this function to the number {\tt 5}:

\[
( \lambda x:{\tt nat}. {\tt 1} ) {\tt 5}
\]

Usually, such a scenario is captured by a typing environment $\Gamma$
(also called context) with variables and their types.  Formally, a
typing environment (or context) is a set of pairs $x:\tau$ where $x$
is a variable and $\tau$ is its type.  For example, if we only have
base variables at the beginning of execution, then the typing
environment for the previous example would be
$\Gamma = \{ \tt 1 : \tt nat, \tt 2 : \tt nat, ..., \}$.
Traditionally, a type judgment is denoted as follows:

\[
\Gamma \vdash e : \tau
\]
which means that the term $e$ has type $\tau$ in context $\Gamma$.
Note that going forward, we will refer to the typing environment
simply as the context.



\subsection{Type Inference}

Type inference algorithms compute the type of an expression.  Type
inference plays an important role in typed programming languages but
becomes more difficult as languages become more expressive; for
example, type inference in OCaml or Scala is EXPTIME (but almost
always tractable in practice) \cite{Kfoury1990, Mairson1990}. 
In more expressive type theories, such
as those used for theorem proving, type inference becomes undecidable
in general and difficult to automate even in practice with extensive
use of heuristics.

Type inference in the simply typed lambda calculus, however, is trivial.
In particular, its PTIME-completeness is well known \cite{Mairson} (i.e., one can compute the type in polynomial time using a Turing Machine) and to show its triviality provide the pseudocode in section \ref{simply_typed_lambda_calc_tinfer}.
The triviality of this problem
is exactly why we hypothesize that type inference for simply typed
lambda calculus is an excellent few-shot learning benchmark.  Given a
few examples of type inference in a programming languages class,
students should be able to generalize to terms of longer length, terms
with different base types, etc.  A transformer that understands the
formal language model over which it is operating should be able to
perform type inference on the simply typed lambda calculus.

\section{Data set for Type Inference}

Recall that we train a transformer to explore arguably the most
fundamental relationship between terms and types: can a model infer
the type of a simply typed lambda calculus term?  (i.e., learning to
do type inference).
In particular, we train a transformer on a
synthetic data set we generated.  In this section, we describe the
synthetic data set generation with some concrete data examples.

\subsection{Synthetic Data Set Generation}

When generating a synthetic data set of the simply typed lambda
calculus, we need to have a typing environment $\Gamma$ with base
variables and base types as explained in section
\ref{lambda_calc_background}.  In our experiments, we had the initial
context with a single vase variable $\tt x$ with base type $\tt T$,
giving the context
\begin{equation}
\label{eq:Context}
\Gamma = \{ \tt x : \tt T \}
\end{equation}

\textbf{Dataset generation:} To generate a single instance in the
dataset, we generate a random type $\tau$ and then proceed to generate
a term $e$ with that type. 
This process is trivial in simply typed lambda calculus because one only needs to inspect the constructors of the Abstract Syntax Tree (AST) of the term, but it is undecidable in general \cite{Chlipala2019}.

In addition, we have all terms to be bounded according to the depth of
their AST.



\textbf{Random type generation: }
  \textit{Input:} To generate types we need: 
  a list of valid base types from the global context $\Gamma$, 
  the maximum allowed depth for any type and 
  the probability distribution for when to generate a base type or to branch  and thus continue recursively generating a type.
  \textit{Output:} A list of types. 
  Note that a data set might have repeated types, but we do not allow an intersection of repeated types between the train, validation and test sets.



When recursively generating a type, we first flip an unbiased coin whether we should branch (and thus recursively keep building a deeper term) or return a base type.
If our coin flip suggests branching and recursively keep building a tree, we first check if we have reached the maximum depth - if we have, we instead return a base type instead of recursively building a deeper term.
If we choose to branch, then it means we will create an application type and proceed
to generate the left and right part of the arrow type recursively as described.
Note that we can do not always generate complete binary tries because we have some probability of continuing deepening the term or stopping to return for that subtree a base type.

\textbf{Random term generation: } 
\textit{Input:} To generate a random term, we receive as input the desired type $\tau$, the (current) dynamic context $\gamma'$ (which starts off as the global context $\Gamma$), the probability distribution for branching and the maximum allowed depth for any term. 
\textit{Output: } We output a single term $e$ of this type $\tau$.
To generate a term, we begin by inspecting what sort of type $\tau$ is and accordingly (randomly) build a term of that type.
Note that every time we generate an abstraction, we make sure to add the bounded variable to the context so that the right context is always available.
We call this the (current) dynamic context $\Gamma'$, and it starts being equal to the global context with the base variables and their types.
Our implementation makes sure that binding rules apply as normal when there are name clashes for nested abstractions.
For example, if there are two variables named {\tt y }, then an abstraction closer to the root would be overwritten by an abstraction created later (deeper) in the lambda term.
When generating a term, there are two main cases to consider: 

1. if the given type $\tau$ is a base type, or 

2. if $\tau$ is an arrow type $\tau' \to \tau''$ (for an abstraction).

In the first case (when $\tau$ is base type), then we have two options, 

a. to generate a term by choosing a variable from the current dynamic context $\Gamma'$ with that type, or

b. (recursively) generate an application - of the form $e_{left} (e_{right})$ - that would eventually return the desired type $\tau$ when evaluated. 
To do this we generate a new type $\tau' \to \tau$ for $e_{left}$ and then recursively generate a $e_{left}$ term with this type.
Then we recursively generate a term $e_{right}$ with type $\tau'$.
Given those two, we return an application term of the form $e_{left} (e_{right})$.
For case 1, we decide to branch according to 1.a. or 1.b. with uniform probability and if at any point we reach the maximum depth, then we do 1.a. to avoid making the term deeper.


Now we proceed to describe how to take care of case 2. - when we need to generate a term of type $\tau' \to \tau''$.
In this case, we have four options when generating a term of such type:

a. we can choose a variable in the dynamic context $\Gamma'$ of that type - if there is one (e.g., it could have been added when creating an abstraction).

b. we choose to create a new abstraction that directly has that type $\tau' \to \tau''$

c. we choose to create an application that if evaluated it ultimately returns a term of type $\tau' \to \tau''$

d. the current depth of the term has reached the limit - so we choose to return a variable from the dynamic context or an abstraction of depth 1.

We choose options 2.a., 2.b., or 2.c. uniformly at random.
Step 2.c. proceeds identically to step 1.b. except that the overall term returned is of type $\tau' \to \tau''$.

To create an abstraction for step 2.b. we need to create a bound variable and its body.
The bound variable is created such that we do not reuse a name from the base variables - if this were to happen, it would be like having a name like {\tt 1} disappear from the global context, which would be fatal.
Besides that, we can create any string that has already been used or a fresh variable with uniform probability over all these valid names.
To create its body, we simply create a term recursively using the procedure outlined in this section, with the input type being $\tau''$.

 

We proceed to give a few concrete examples of possible data points
$(x, y)$, where $x$ is the term and the target type to infer is $y$ in
context $\Gamma = \{ \tt x : \tt T \}$.  Example 1:
\[
(x, y) = (\tt x, \tt T) 
\]
In this example, the term is the base variable $\tt x$ with base type
$\tt T$.

Example 2:
\[
(x, y) = 
({\tt x_0 : T . x},
{\tt T \to T})
\]
This example shows a lambda abstraction with bound variable $\tt x_0$
with type $\tt T$.  The lambda abstraction returns the base variable
$\tt x$ for any term of type $\tt T$.  Thus, the overall type is
$\tt{T \to T}$.

Example 3:
\[
(x, y) = 
(\tt{ (\lambda . x_0 : T. x) x}, 
\tt{T}) 
\]
In this second example, the term is the application of the lambda
abstraction from the previous example - applied to the base variable
$\tt x$.  Since this lambda abstraction always returns the base
variable $\tt x$ for any argument, it always has the $\tt T$.

Example 4:
\[ 
(x, y) =
(\tt{\lambda x_0 : T \to T} .
\tt{ x },
\tt{(T \to T) \to T} )
\]

This example shows a lambda abstraction with bound variable $\tt x_0$ with type
$\tt T \to T$.  In particular, notice that the argument is explicitly
a function type from the base type to the base type $\tt T$.  Since
the body always returns the base variable $\tt x$, the resulting type
is from the input type $\tt T \to T$ to the output type $\tt T$.

Example 5:

\[ 
(x, y) =
(\tt{ 
(\lambda x_1:T.(\lambda x_2:T \to T . x_2) x) (\lambda x_0:T.x_0)
} ,
\tt{T \to T} )
\]
This is a slightly more complex data example, where we apply the
lambda abstraction
$\tt{ (\lambda x_1:T.(\lambda x_2:T \to T . x_2) x) }$ to the
argument $\tt{ (\lambda x_0:T.x_0) }$.  Observe, however, that the
term $\tt{ \lambda x_1:T.(\lambda x_2:T \to T . x_2) x}$ evaluates
to $\tt{ \lambda x_2:T \to T . x_2 }$ for any argument of type
$\tt T$.  
Therefore, overall, example 5 applies the function
$\tt{ \lambda x_2:T \to T . x_2 }$ to the argument
$\tt{ (\lambda x_0:T.x_0) }$ which results with a type
$\tt{T \to T}$.

As a final remark, we want to remind the reader that the initial global environment was $\Gamma = \{ {\tt x} : {\tt T} \}$ - so all our simply typed lambda terms have a single base type ${\tt T}$.

\section{Model Architecture}

Our model consists of a standard transformer \cite{Vaswani2017} model that predicts rules from a grammar to generate types for simply typed lambda calculus terms. 
Motivated by the rich syntactical structure information contained in formal languages, our model takes in a preprocessed sequence representation of the CST of the inputs, instead of the raw sequence of tokens from the text.
Since we are predicting the type given a term, the encoder takes in the preprocessed sequence representation of the term and the decoder takes in the standard right-shifted preprocessed sequence representation of the type.
The model had an embedding dimension of size $1024$, the transformer had $3$ layers for both the decoder and encoder.



\subsection{Preprocessing of names and the open vocabulary problem}

Before producing a sequence representation of the CST, we preprocess the name inside the terms as described in this section.
First, we motivate this preprocessing.
In program synthesis there does not exist a closed vocabulary as in natural language.  
For example, programs can have arbitrary names for variables, function, types, or any object.
To avoid having to solve this challenging open problem in our experiments, we assume there is a fixed global context that allows us to know in advances all the symbols that can be seen by our model.  
Therefore, we only use $32$ different bound variable names to keep our closed vocabulary assumption true.  
In addition, all bound variables have been preprocessed such that they are renamed according to their Breadth-First Search (BFS) ordering in the CST.  
Therefore, we modify the grammar before training (and inference) and replace the regular expressions in our grammar (in section \ref{grammar}) with the names from the global context - namely $\Gamma = \{ \tt{x:T} \}$ - and the new unique $32$ BFS variable names.  
With this machinery in hand, we proceed to explain how we create the input CSTs to the encoder and decoder of our model.

\subsubsection{The CST input to the Encoder and Decoder }\label{encoder_input}

\textbf{Input to Encoder:}
The encoder takes in a sequence of tokens from the BFS traversal of the CST from a simply typed lambda calculus term.
To produce a sequence of tokens, we traverse the CST of the term in BFS order using the edge symbols and node symbols.
The node symbols are non-terminals from the grammar for intermediate nodes, and terminal symbols for the leaf nodes.
In addition, we append the special tokens ``sos", ``eos" at the start and end of the sequence for each term in a batch.
Finally, we add the positional embedding from the original transformer \cite{Vaswani2017} to the input sequence.
In an input batch to the encoder, the terms might have different lengths.
To address this, we append a pad token to the shorter term sequences up to the longest term sequence in a batch.
During training and inference, these pad tokens are masked and never affect the output of the model.
Note that a traversal ordering (e.g., BFS for us) of the sequence is required if one wants to use an autoregressive model (like the transformer) that takes in a sequence as input - instead of a more structured input like a CST.

\textbf{Input to Decoder:}
We follow the same input format to the decoder as in the standard transformer \cite{Vaswani2017}.
In particular, since it is trying to predict a sequence of grammar rules for the type - it therefore takes in the right-shifted sequence of embeddings for grammar rules.
To generate this input, we produce a sequence of grammar rules from the BFS traversal of the CST of the target type.
Note that a traversal like this is forcibly required when transforming a structured input like a CST into a sequence compatible with an autoregressive model.
Then we right-shift the sequence according to a single token and then append the special tokens ``sos", ``eos" at the start and end of the sequence for each target type in a batch.
Similarly, as in the encoder input, we append the pad token to address the variable length of types and mask accordingly to avoid the output of the model from being affected.
We also apply the positional embedding from the original transformer \cite{Vaswani2017} to the input sequence.

\textbf{Remarks on the presence of types in the input term:} 
We would like to emphasize that the types for the bound variables of the lambda abstractions are not hidden from the model.
This information is essential for the task of type inference to not be ill-posed.
Without this information it is impossible to solve the type inference problem with further distributional assumptions, e.g., our data comes from terms produced by humans.
Therefore, types are present in the input term to the encoder.


\subsection{Encoder and Decoder stacks}
We use the PyTorch API \cite{pytorch} for our transformer's encoder and decoder layers, and thus all layers are present exactly as described by the original authors \cite{Vaswani2017}.
We depict this abstraction in figure \ref{our_model} and chose to only make explicit the positional encoding - because of different ablation experiments we do.
For further details, we refer the readers to the original transformer paper \cite{Vaswani2017} and the PyTorch API \cite{pytorch}.
The hyperparameters used for results in tables and figures were: $1024$ for all embedding dimensions, 
$3$ for the number of layers for both the encoder layer and decoder layer, 
and $0.1$ dropout rate.
The hyperparameters used for results in tables and figures were: $256$ for all embedding dimensions, 
$1$ for the number of layers for both the encoder layer and decoder layer, 
and $0.1$ dropout rate.

\subsection{Grammar Rule Prediction}

We predict the ID of grammar rules of the types from the lambda
calculus after the type grammar in equation \ref{eq:grammartype} has been preprocessed with the allowed base types from the global context.
For us this global context is $\Gamma = \{ \tt{x:T} \}$.
Note that in practice the two grammars are present in a single grammar as in section \ref{grammar} - since the terms are simply typed the term grammar contains the type grammar.
To predict the grammar rules, we follow the information flow depicted in figure
\ref{our_model} in our model and use the decoder's last feature layer
as input to a linear layer that predicts the ID of the grammar rule
for types.

\subsection{Training}\label{training}

To train our model, we convert the target type into a rule sequence in BFS
ordered.
We compute the loss by computing the cross-entropy loss of the model's predicted rule sequence and the rule sequence of the target type. 
Recall that the focus of this work is the study of the different behaviors of optimizers with a wide variety of hyperparameters.  
Therefore, the optimizers we used are: Adam, RAdam \cite{radam} and Adafactor \cite{adafactor}.  
The experiment section \ref{expts} outlines the different ablation studies and hyperparameters we use for our training experiments.

\subsection{Inference - Type Synthesis}\label{synthesis}

For the synthesis of types, we go through the model's predicted
distribution of rule sequences and synthesis a type greedily by
choosing the ID of the most likely grammar rule.  When we synthesize
types, we expand the model's predicted rules in BFS order, since the
model was trained with sequences ordered in this way.  If at some
point during synthesis, a pair of rules in the predicted grammar
sequence could not have been applied in BFS order - we predict the
dummy error type.  For more details, refer to algorithm 1.

\begin{algorithm} 
	\caption{Greedy Synthesis of Types in BFS order} 
	\begin{algorithmic}[1]
	    \State{\textbf{Input: } batch of terms $B$ to synthesize types of length $|B|$}
	    \State $\tau = [] $ initialize the empty list to hold the predicted batch of types.
		\For {$b = 1,2,...,|B|$}
		    \State $T_b = $ get a term from a batch of terms $B$ with index $b$
		    \State $R_b = $ get the predicted rule sequence $R_b$ for the term $T_b$
		    \State $G_b = [] $ initialize an empty list to hold the greedy rules from $R_b$ 
			\For {$t = 1,2, ..., |R_b|$}
			    \State $p_{b,i}$ = get distribution of possible rules for the current type inference step $t$
			    \State $\hat r_{b, t} = arg \max_{r \in {1,2,...,|p_{b,i}|} } \{ p_{b,t}[r] \}$ i.e., get the most likely rule for this time step $t$ for the current term $T_b$
			    \State $G_b[t] = r_{b, t}$ i.e., store the greedy rule to predict in time step $t$
			\EndFor
			\State $\tau_b = $ generate type by applying the greedy rules from $G_b$ in BFS order. 
			If a rule cannot be applied with respect to the previous rule in BFS order skip this type and append the dummy error type.
			\State $\tau[b] = \tau_b$ append the predicted type $\tau_b$ to the list of types.
		\EndFor
		\State{\textbf{Return: } batch of predicted types for input terms $\tau$}
	\end{algorithmic} 
\end{algorithm}
\begin{figure}[h]
\includegraphics[width=1.0\linewidth]
                   {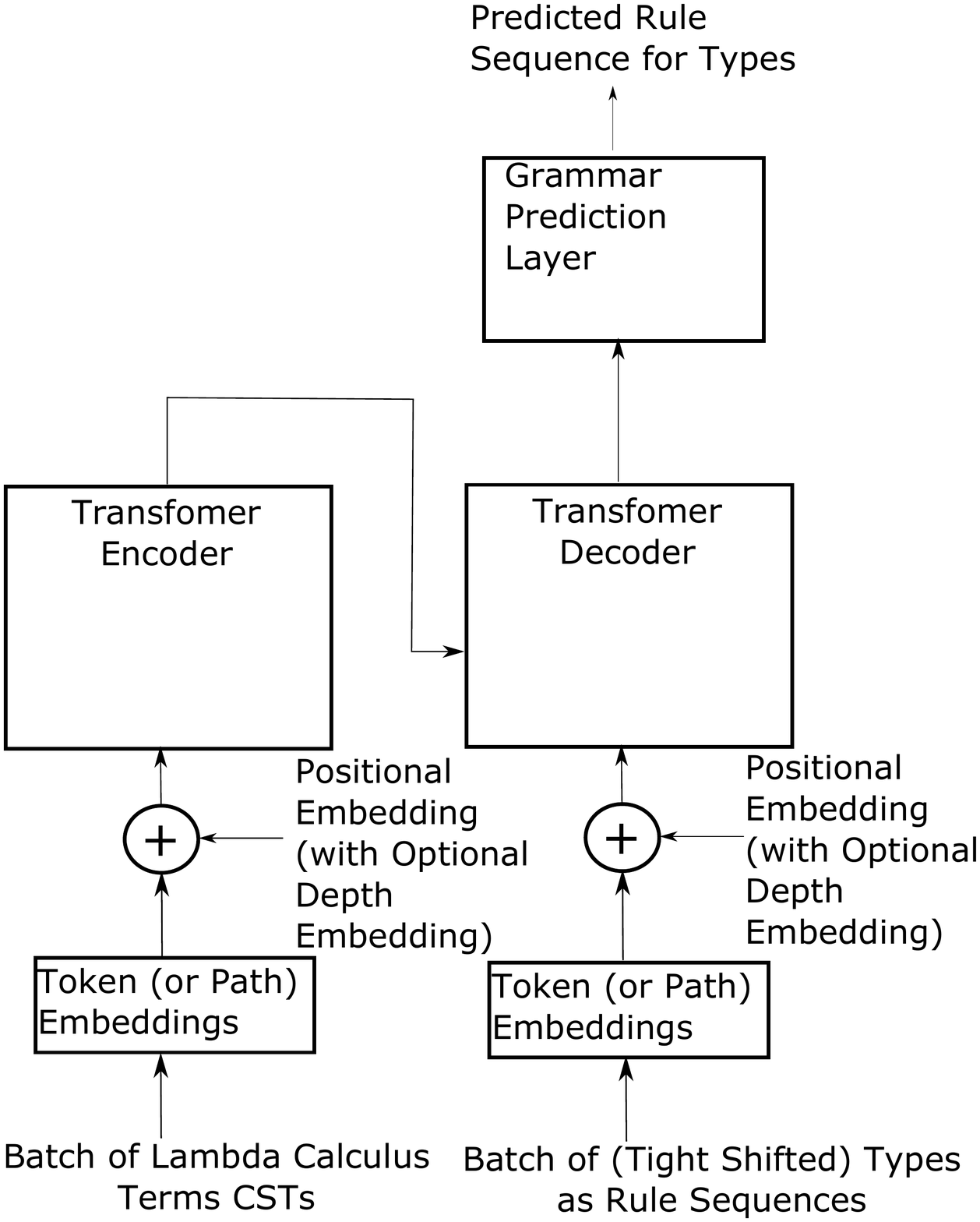}
\caption{Our transformer model for simply typed lambda calculus.}
\centering
\label{our_model}
\end{figure}

\section{Experiments}\label{expts}

In this section, we present the evidence to support our conclusion that we solved the extensive hyperparameter search problem for training
transformers to solve the type inference problem - by using the Adafactor optimizer.
  

\subsection{Is the warm-up phase important for transformers for learning to predict simple types?}

The literature on transformers has a strong emphasis on the importance
of choosing an appropriate warm-up phase for stable learning
\cite{trans_tutorial, Vaswani2017, radam,
  on_the_adequacy_of_untuned_warmup, adafactor}.  

The warm-up phase is
an initial period of training where the learning rate starts at zero
and is increased to a target value.  In this section, we explore the
importance of the number of warm-up steps with various learning rates
using Adam \cite{adam}.  Main results for this subsection are in table
\ref{table1}.

\begin{table}[h]
\begin{center}
\begin{tabular}{|c|c|c|c|}
\hline
 & \multicolumn{3}{|c|}{ Train Accuracy } \\
\hline
 number warm-up steps & lr=$10^{-3}$ & lr=$10^{-4}$ & lr=$10^{-5}$ \\
\hline
0 & 0.31 & 0.95 & 0.99 \\
2,000 & 0.089 & 0.93 &  0.99 \\
4,000 & 0.27 & 0.76 &  1.0 \\
10,000 & 0.31 & 0.97 &  0.99 \\
\hline
\end{tabular}
\end{center}
\caption{ 
Results showing the role of the number of warm-up steps on the training accuracy.  
We train our transformer model shown in figure \ref{our_model} with token embeddings from CSTs with only positional encoding - as described in section \ref{encoder_input}.  
Our model was trained with batch size $128$ for $80,000$ iterations for $2$ days and $4$ hours.  
Our model had an embedding dimension of size $1024$, the transformer had $3$ encoder and decoder layers.  }
\label{table1}
\end{table}

From table \ref{table1} we can see that as the number of warm-up steps
is varied, the training accuracy of the transformer does not change.
In particular, for a learning rate of $10^{-5}$ we notice the model
stays at perfect prediction no matter the number of warm-up steps.
For a learning rate of size, $10^{-4}$ we can see that the model
achieves good performance on most runs but not perfect accuracy
compared with the $10^{-5}$ - suggesting that a good choice of
learning rate is more important than having a warm-up phase.  For the
largest learning rate of size, $10^{-3}$ one would expect the biggest
benefit from a warm-up phase, but the performance of the model gets
capped at $0.31$ - even when having a long warm up phase of $10,000$
steps.  Note that $2,000$ iterations of warm up is in line with the
suggestions from previous work
\cite{on_the_adequacy_of_untuned_warmup} - suggestion a warm-up of
$2 (1 - \beta_2)^{-1}$ (where $\beta_2 = 0.999$ is the hyperparameter
of the second moment from the Adam optimizer \cite{adam}).

These experiments suggest that choosing a good learning rate seems
sufficient for a transformer to learn to do type inference.

\subsection{Does RAdam fix the convergence issues?}

Motivated by the big emphasis in the literature on a warm-up phase
\cite{radam, trans_tutorial, Vaswani2017,
  on_the_adequacy_of_untuned_warmup} and complexity of tuning warm-up
hyperparameters we tested Rectified Adam (RAdam) - an optimizer that
claims to fix the high variance problem of Adam at the beginning of
training \cite{radam}.

\begin{table}[h]
\begin{center}
\begin{tabular}{|c|c|c|c|}
\hline
 & \multicolumn{3}{|c|}{ Train Accuracy } \\
\hline
 number of warm up steps & lr=$10^{-3}$ & lr=$10^{-4}$ & lr=$10^{-5}$ \\
\hline
0 & training diverged & 0.91 & 1.0 \\
2,000 & - & 0.96 &  0.99 \\
4,000 & - & 0.80 &  1.0 \\
10,000 & - & 0.96 &  0.99 \\
\hline
\end{tabular}
\end{center}
\caption{ Results showing the role of number of warm-up steps on the
  training accuracy with RAdam \cite{radam}.
  We train our transformer model shown in figure \ref{our_model} with token embeddings from CSTs with only positional encoding - as described in section \ref{encoder_input}. 
  The model was trained with batch size $128$ for $80,000$ iterations for $2$ days and $4$ hours.  
  The model had an embedding dimension of size $1024$, the transformer had $3$
  layers for both the decoder and encoder. 
  The table entries for a learning rate of $10^{-3}$ with warm up steps of $2000, 4000$ and
  $10,000$ were not run.
  We choose to do this since the warm-up would increase the learning rate and would likely lead to similar training divergence.  }
\label{table2}
\end{table}

From table \ref{table2} we see that RAdam behaves similarly to Adam
with or without a warm-up phase.  This is not surprising because RAdam
is supposed to fix the initial instability of training.  However, the
evidence from table \ref{table1} suggest that a warm-up phase does not
provide a bit of stability of our task.  This suggests that RAdam
would not have a big impact either.  This is highlighted by the fact
that when no warm-up schedule is used (row 1 of table \ref{table2}) -
the model diverges and then the model fails to get perfect accuracy
until the right learning rate is found.

We did not try in depth experiments with $10^{-3}$ because the
consistently training diverged for us.  To address this issue, we did
try decaying the learning rate with two different decay schedulers
with RAdam.  We tried a decay scheduler (with no warm-up) that decayed
when the training loss was on a plateau, and this did not help
convergence.  In another experiment, we used the exponential decay,
decaying every epoch.  This did not help either.  Neither of the
approaches solved the divergence issues for a learning rate of
$10^{-3}$.  But further experiments would be interesting.

We would like to emphasize that in addition to choosing the learning
rate when training transformers, one also has to choose the number of
steps to train Adam or RAdam as an additional hyperparameter choice.
In addition, this hyperparameter interacts with the number of warm-up
steps - complicating the choice of both
hyperparameters. 

\subsection{Experiments with Adafactor}

Motivated by the difficulty to train our model to achieve perfect
train accuracy by Adam and RAdam on a small dataset,
we tested Adafactor \cite{adafactor, adafactor_huggingface} on our task.  To check the robustness of
Adafactor, we also provide additional experiments with different
variants of the model in table \ref{table3}.  
We describe these variants in the appendix section \ref{depth_path_embedings}.

The training succeeded in all these settings and reached perfect train accuracy within 2-3
hours with smooth learning curves - similar to the ones shown in
\ref{learning_curves}.  To see the effect of an annealing scheduler on
the learning rate, we used the default Adafactor scheduler and found
that this allowed a 2.5-fold speed up to reach the same train
accuracy.  When calling the scheduler every epoch, the training time
decreased from 10 hours to 2-4 hours.

\begin{table}[H]
\tabcolsep=2.2pt\relax 
\begin{center}
\begin{tabular}{|c|c|c|c|c|}
  \hline
  embedding method & train accuracy & validation accuracy & iter/sec & runtime \\
  \hline
  token (nlp) & 1.0 & 0.99 & 0.2 & 1.7 hours \\
  path + $f_{char}$ & 1.0 & 0.99 & 0.2 &  2.1 hours \\
  Depth Embed + $DE_{RI}$ & 1.0 & 1.0 & 0.2 &  2.4 hours \\
  \hline
\end{tabular}
\end{center}
\caption{ Results showing the success of training a model with
  Adafactor \cite{adafactor} with different methods to embed the
  lambda calculus terms.  We choose the setting as outlined in the
  documentation for the Hugging Face Adafactor
  \cite{adafactor_huggingface} with a scheduler called every epoch.
  In this case, it means we called it every $78$ iterations for our
  task.  Note, that the models also converge without a scheduler, but
  it took about 10 hours instead of about 2h.  We trained the
  transformer model as described in figure \ref{our_model} with
  different embeddings as outlined in the first column.  The first row
  (nlp) refers to a pure token model as used in NLP described in
  section \ref{encoder_input}.  The second row (path+$f_{char}$) denotes a
  model that embeds the path as described in section \ref{depth_path_embedings}.  The
  third row (Depth Embed + $DE_{RI}$) denotes a depth embedding
  according to the parent grammar rule that generated the token, as
  described in section \ref{depth_path_embedings}.  The model was trained with a batch
  size $128$ until it reached perfect training accuracy.  The model
  had an embedding dimension of size $1024$, the transformer had $3$
  layers for both the decoder and encoder.  }
\label{table3}
\end{table}

We would also like to highlight that \textbf{our model had perfect
  validation accuracy}.  Suggesting that the model generalizes
perfectly when the terms used for training and testing have the same
fixed depth.  However, we conjecture that search methods (like beam
search) will be needed when testing for systematic generalization.
For example, when the terms in the validation set have a larger depth
size than the depth of terms used to train the model.

\subsection{Comparison of Adam's and Adafactor's Learning curves}

\begin{figure}[H]
\includegraphics[width=1.0\linewidth]
                   {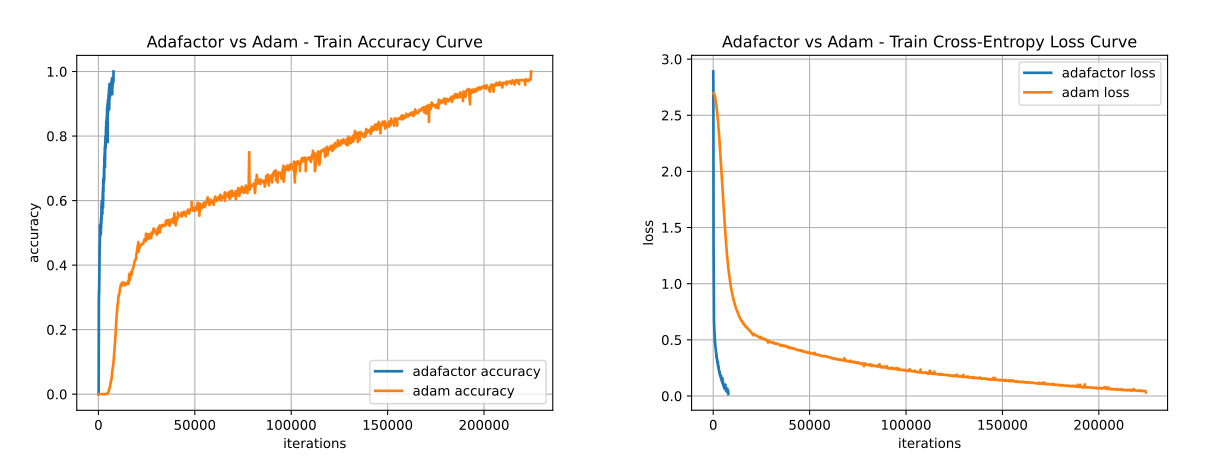}
\centering
\label{learning_curves}
\caption{ {\bf Adafactor reaches perfect train accuracy 11 times faster than Adam.}
Both optimizers were trained until perfect accuracy on the training set was achieved. 
Adafactor used the default scheduler provided by Hugging Face \cite{adafactor_huggingface} and Adam used a linear warm-up schedule.
The learning rate $10^{-5}$ for Adam and Adafactor used the default Hugging Face \cite{adafactor_huggingface} hyperparameters, which included no search for a learning rate for us.
No hyperparameter - except calling the scheduler every epoch - was done for Adafactor.
Adafactor was trained for $7,821$ iterations until it got perfect accuracy with its scheduler being called every two epochs - in this case around $146$ iterations.  
Adam was trained for $224,439$ iterations until it got perfect accuracy in 1 day 15 hours 43 minutes. 
Adafactor was trained for $7,821$ iterations until it got perfect accuracy in 2 hours 10 minutes.
Note that we did try annealing schedules on Adam with no success on beating Adafactor.
}
\end{figure}

Motivated to understand the difficulty of training the transformer
model in figure \ref{our_model}, we analyze the learning of the
successfully trained experiments depicted in figure
\ref{learning_curves}.  The main observation is that Adafactor
\cite{adafactor} with an annealing scheduler is able to train in just
2.17 hours, while it took 1 days and 15 hours to train the same model
with Adam using a linear warm-up phase.  
We chose the annealing
scheduler rate according to the following equation
\[ \mathsf{scheduler\_rate} = \min( \mathsf{one\_epoch}, 2(1 -
  \mathsf{beta\_2})^{-1} ) \]

which resulted in calling in scheduler rate of every epoch.
Note, however, that even without an annealing scheduler, our
transformer could train successfully but took about 10 hours instead.

\subsection{Hardware and framework parameters}
All times are reported with respect to a measurement performed on a
single IBM Research Cluster virtual node using 1 GPU (NVidia Tesla
P100 with 16 GB GPU RAM) and 16 virtual CPU cores (Intel Xeon Platinum
8260 (R) at 2.40GHz with 118 GB CPU RAM).  The implementation used PyTorch
1.9.1 \cite{pytorch}.
We also used the ultimate-utils library \cite{brando2021ultimateutils} -- a utils library for machine learning -- to support our PyTorch model and training implementations. 

\section{Discussion}

In this paper, we solved the extensive hyperparameter search problem
when training transformers to learn type inference by using the
Adafactor optimizer \cite{adafactor}.  This approach has a number of
benefits: a) it can get a 2.5-fold speed up by adding an annealing
scheduler called every
\[ \mathsf{scheduler\_rate} = \min( \mathsf{one\_epoch}, 2(1 -
  \mathsf{beta\_2})^{-1} ) \]

(in this work every epoch) b) the method works even
without a scheduler and one only has to track when the model has
obtained perfect accuracy (or has converged).  In addition, the
original authors of the Adafactor optimizer designed it such that it
uses sublinear memory - an especially valuable feature in the era of
huge language models.  In addition, due to the difficulty to choose
hyperparameters, Adafactor has the ability to speed up the research
process and recommend practitioners to use it without a scheduler out
of the box with transformers.  The choice of Adafactor over Adam is
highlighted, that the hyperparameter search for Adam - in particular
the number of iterations - was guided using experiments that had
already succeeded with Adafactor.  To our surprise, we found RAdam to
diverge more than Adam and do not recommend it for our model and task.

Finally, we conjecture that training on programming language tasks -
like type prediction - has a different optimization landscape compared
to training on natural language.  We hypothesize this is the case
because when one changes a single token of a term, this leads to
completely different semantics, while changing a single token of a
sentence in natural language often only makes the sentence awkward.
Our evidence suggests that, even, training a transformer on the
simplest of tasks - predicting the type of a simply typed lambda
calculus term - requires careful optimization.  Independently, related
work discovered the difficulty of training transformers of formal
rules and called their observation the ``Grokking effect".  They
observed that after long periods of training, the transformer model
can go from random chance to perfect generalization.  Based on our
evidence, we conjecture that Adafactor might be able to resolve this
and suggest it for future work.  In addition, we conjecture that even
with Adafactor, the transformer is unlikely to exhibit systematic
generalization and truly learn the underlying rules and generalize to
predict terms with larger depth than it was trained on.

\section{Related Work}

The incorporation of unique learning schemes to train transformers on
natural language tasks can be observed since the original transformer
\cite{Vaswani2017} which used the following learning rate scheme:
\[
\mathsf{lrate} = 0.0325 \cdot \min( \mathsf{step\_{num}}^{-0.5},
\frac{\mathsf{step\_{num}}}{252982}) \]

This corresponds to increasing the
learning rate linearly and then decreasing it inversely proportional
to the square root of the step number.

Later work investigated the need for the warm-up stage and identified
the high variance in the adaptive learning rate \cite{radam} with
their RAdam optimizer.  However, later work showed that the variance
of the adaptive learning rate, although divergent, does not result in
a divergent update term \cite{on_the_adequacy_of_untuned_warmup}.
This new follow-up work suggests that the number of warm-up steps of
$2 \cdot (1 - \beta_2)^{-1}$.  However, our experiments show that the
number of warm-up steps does not affect the convergence of the model
in our task, but instead the type of optimizer used or using low
learning rate with long training is sufficient.  Our evidence suggests
that the difficulty of training transformer-based models for learning
formal rules - named ``Grokking" by previous work \cite{grokking} -
could be avoidable by using the Adafactor optimizer \cite{adafactor}
and suggest practitioners and researchers to try it first.  Finally,
the Adafactor \cite{adafactor} optimizer was designed with the
objective of having a sublinear memory footprint for transformers, and
our experiments show that this optimizer results in a nice stable
optimizer with only 1 potential hyperparameter to tune.  Future, work
could test Adafactor's stability with a larger set of programming
language challenges for the transformer.

\subsubsection*{Acknowledgments}
We'd like to thank Tejas I. Dhamecha at IBM for sharing
their experience in training Transformer neural networks and
discussions over slack during the summer of 2021.  



\medskip

\bibliographystyle{unsrt}
\bibliography{egbib}

\section{Appendix}

\subsection{Lambda Calculus Grammar}\label{grammar}

Here we present the lambda calculus grammar in EBNF format using the
Lark Python syntax used in our code.  
Recall that this grammar gets
expanded into a closed vocabulary by appending 32 bound variables in
addition to the global context.
White spaces and parenthesis are ignored as tokens in accordance to
our grammar. 
The result grammar results in the following\footnote{
  \url{https://github.com/FormalML/type-parametric-synthesis/blob/master/tinfer/simply_type_lambda_calc.ebnf}}
\begin{verbatim}
term : "lambda" term ":" type "." term
    | "[" term term "]"
    | /[A-Za-z0-9_]+/
type : type "->" type  // "->"
    | /[A-Za-z0-9_]+/
IGNORE_TOKENS : "(" | ")"
%import common.WS
%ignore WS
%ignore IGNORE_TOKENS
\end{verbatim}

Note that the grammar for terms and types is intertwined because the terms have type annotations.

\section{Type inference in Simply Typed Lambda calculus}\label{simply_typed_lambda_calc_tinfer}

To infer the type of a simply typed lambda calculus term $e$ with the
context $\Gamma$, one traverses the AST of term and builds the type
recursively according to the constructor sort at each level of the AST
of $e$.
Overall, the process for type inference is outlined in algorithm 2.

\begin{algorithm} 
	\caption{Type inference for Simply Typed Lambda Calculus} 
	\begin{algorithmic}[1]
	    \State{\textbf{Definition }infer\_type($e$, $\Gamma$)}:
		\State{\textbf{Input:} 
		$e$ input term we want to infer the type. 
		Context $\Gamma$ with term-type pairs $x:\tau$.}
		\If{ term $e$ is a $Variable$}
		    \State{ $\tau_e = \Gamma.get\_type(e)$ }
		    \State{ \textbf{Return:} $\tau_e$ }
		\ElsIf{term $e = \lambda x: \tau. body$ is an $Abstraction$} 
			\State{ $\Gamma' = \Gamma \cup \{ x: \tau \}$ }
			\State{ $\tau_{abs} = infer\_type(body, \Gamma' ) $ }
		    \State{ \textbf{Return:} $\tau_{abs}$ }
		\Else{ term $e = (e_{left}) e_{right}$ is an $Application$ \textbf{then}} 
			\State{ $\tau_{left} = infer\_type(e_{left}, \Gamma) $ }
			\State{ $\tau_{right} = infer\_type(e_{right}, \Gamma) $ }
			\State{ $\tau_{apply} = \tau_{left} \to \tau_{right}$ }
		    \State{ \textbf{Return:} $\tau_{apply}$ }
		\EndIf
	\end{algorithmic} 
\end{algorithm}







\section{Variations on Positional Encoding for CSTs}\label{depth_path_embedings}

To test the robustness of Adafactor we also augment the initial input sequence form the CSTs with additional structural information - similarly to how the TreeGen model \cite{treegen} does.
We do this in two ways: 
1. by providing an embedding of tokens for the CST based on the path to each token in the CST and,
2. by using a depth encoding layer in addition to a positional encoding layer
We review these methods, provide the motivation for them, and clarify our adaptation for our robustness experiments.

\subsection{Motivation}

Our initial experiments revealed the unexpected challenge to train transformers to learn to do type inference.
We discovered that the Adafactor optimizer was a surprisingly effective strategy to train our model.
To test if this was true in general in our setting, we decided to test our model with Adafactor but incorporate variations in the way to embed the input model - to test the robustness of Adafactor.
Table \ref{table3} of our main results show that indeed the model remained robust, but there were no improvement in the convergence time.

\subsubsection{Variations to the Encoder and Decoder inputs}\label{encoder_input}


\textbf{Path embedding method:}\label{path} 
Motivated by the fact that using BFS ordering of symbols to produce sequences loses structural information about the term, we implement path embeddings similar to \cite{treegen}.
This method replaces each node in the CST with the path to that node -
to encode this structural information in the tokenization of the input. 
A path is the sequence of symbols visited when traversing to a specific node in the CST 
- including the symbol of the final visited node.  
Therefore, we embed tokens using the paths to their nodes using a feed forward neural network - similar to the one described in \cite{treegen}.
The path embedding computation is
\begin{equation}
  \mathsf{embed}(x_{\mathrm{path}}) := \mathsf{FFD} [x_1; \dots; x_L  ]
\end{equation}
where $\mathsf{FFD}$ is feed-forward network with trainable weights 
where $L$ is the maximum length of a
path, $x_i$ is the embedding of the symbol in the path 
  The $\mathsf{FFD}$ outputs path embeding in $R^{M}$
 Here by $[x_1; ...; x_L]$ we denote the
row-wise concatenation operation for character vector embeddings.  In particular,
given a list of $L$ column vectors, the $;$ concatenation operation
stacks them. Therefore, if $x_i \in R^D$ then the matrix $[x_1; ...; x_L]$ is in
$R^{L \times D}$.





We set $L=13$ in our experiments since a path of
length $13$ allows for a reasonably interesting terms - in the case
of binary trees it allows terms with about $8,192$ nodes.  For paths
shorter than $L$, we pad them with a special ``pad" token which
eventually gets masked and doesn't affect the prediction of the model.
Note that path embeddings are analogous to using character embeddings
in NLP.
In NLP, they replace the tokens with the individual characters that compose that token and then embed it.
Analogously, we replace each token with the path to the node corresponding to that token.


\textbf{Depth embedding method:}\label{de} 
Unfortunately, the path embedding method is expensive since it embeds each token using a
sequence of symbols - analogous to character embeddings for words in natural language.  
Since it is not always easy to fully vectorize the code for path embeddings, we instead experiment with depth embeddings as a way to inject structural information into the input sequence (similar to \cite{treegen}). 

Assume we have length $L$ the BST sorted sequence of embeddings vectors for a CST 
$x_{cst\_seq} = [x_{token_1}, \dots, x_{token_L}]$ with embedding size $D$.
We have $x_{cst\_seq} \in R^{L, D}$.
Then the depth embedding $DE$ for this sequence is the embedding sequence using the parent of each token in the CST.
Therefore, $DE_{x_{cst\_seq}} = [DE_{token_1}, \dots, DE_{token_L}] \in R^{L, D}$.
Then we proceed to add this sequence vector to $x_{cst\_seq}$ - the same way one would add the positional embedding:
\begin{equation}
  x_{seq} = [x_{token_1}, \dots, x_{token_L}] + [DE_{token_1}, \dots, DE_{token_L}]
\end{equation}
where $x_{token_i} \in R^D$ is the embedding of the $token_i$ and $DE_{token_1} \in R^D$ is the embedding of the parent of $token_i$ in the CST.
In addition, we always add the usual positional encoding proposed in the original transformer model \cite{Vaswani2017}.

\end{document}